# HOW EVALUATION CHOICES CHANGE THE MEASURED BENEFIT OF COOPERATIVE PERCEPTION: EVIDENCE FROM THREE V2X BENCHMARKS

Pincan ZHAO [a], Yili TANG [a,b,c] and Xinrui ZHANG [d]
[a] Department of Electrical and Computer Engineering, Western University, Canada
[b] Department of Civil and Environmental Engineering, Western University, Canada
[c] Department of Civil Engineering and Building Engineering, University of Sherbrooke, Canada
[d] Department of Computer Science, Lakehead University, Canada
Emails: pincan.zhao@uwo.ca; yili.tang@usherbrooke.ca; xrzhang@lakeheadu.ca

## ABSTRACT

Cooperative perception, in which connected vehicles and roadside infrastructure share sensor information, is a candidate enabler of automated mobility, and benchmark accuracy is the evidence cited when roadside deployment is considered. This paper audits that evidence base across one simulated and two real-world vehicle-to-everything (V2X) benchmarks. In simulation, two widely studied robustness axes leave almost no recoverable headroom: an infrastructure-anchored pose correction returns about one accuracy point at every error level, and corrupting a partner costs 0.7 points. On real data, measurement choices govern the conclusion. An apparent seventeen-fold advantage of infrastructure in partner-poor frames falls below four-fold once the split is broadened, and an ad-hoc class definition measures a far smaller benefit than the official protocol reports. Cooperation is worth 7 to 15 accuracy points, yet 27% of frames on one split offer no partner and almost none do on another, so every regime-conditioned claim must name its split.

## 1. INTRODUCTION

Connected and automated vehicles can exchange sensor information with one another and with roadside infrastructure. A growing body of work shows that such cooperative perception detects road users beyond the line of sight of any single vehicle (Wang *et al.*, 2020; Xu *et al.*, 2022b). Public money is following. The United States national V2X plan reports about USD 200 million of active grant-funded projects, targets V2X capability at a quarter of the signalised intersections in the 75 largest metropolitan areas by 2028, and commits to benefit-cost studies of which deployments are worthwhile (US Department of Transportation, 2024). Siting of individual units is then decided with models that take the communication or perception benefit of a unit as an input (Zeng and He, 2024). Benchmark accuracy is the main published evidence on what a roadside unit contributes, so the reliability of benchmark conclusions is a transportation question and not only a computer-vision one.

That evidence base rests on two foundations, and neither is often audited. Most methodological progress is demonstrated on simulated benchmarks whose traffic and sensing assumptions go unexamined, and evaluation protocols differ between studies in their object classes, detection ranges and test-set composition. Should conclusions move under different but equally defensible measurement choices, method rankings and any deployment argument built on them inherit that fragility. The consequences fall hardest on the regime early deployment occupies: connected-vehicle penetration measured on a large real corridor averages about 6% (Kandiboina *et al.*, 2024), which leaves an equipped vehicle with few partners or none for most of its journey.

This paper reports a systematic empirical audit of that practice. One detection pipeline family is used throughout. We probe a simulated benchmark, V2X-Sim (Li *et al.*, 2022), together with two real-world benchmarks, DAIR-V2X (Yu *et al.*, 2022) and V2X-Real (Xiang *et al.*, 2024). We propose no fusion method. What we ask is what these benchmarks license a reader to conclude, and the audit includes the places where our own earlier measurements misled us.

The contributions are threefold and correspond to Sections 4 to 6. First, two widely studied robustness axes in simulation, pose error and adversarial feature corruption, leave almost no headroom for the remedies we probe. We attribute the closure to dense partner placement and exact infrastructure poses, without isolating it by a control. Second, measurement choices govern conclusions on real data. An apparent seventeen-fold advantage of infrastructure in partner-poor frames falls below four-fold once the data split is broadened, and the cooperation benefit measured under an ad-hoc class definition and range is far below what the official protocol reports. Third, cooperation is nevertheless worth 7 to 15 accuracy points on real data, while partner availability varies sharply between splits. Five recommendations close the paper.

## 2. RELATED WORK

### 2.1 Cooperative perception methods

Cooperative detectors exchange raw sensor data, intermediate features, or final detections, and intermediate fusion dominates recent work. V2VNet propagates features over a spatial graph (Wang *et al.*, 2020), F-Cooper aggregates them by max pooling (Chen *et al.*, 2019), OPV2V uses per-location attention (Xu *et al.*, 2022a), and V2X-ViT applies a heterogeneity-aware transformer that remains the strongest published baseline on real V2X data (Xu *et al.*, 2022b). A second line hardens cooperation against pose misalignment (Lu *et al.*, 2023) and bandwidth limits (Hu *et al.*, 2022). Most of these detectors share the PointPillars encoder (Lang *et al.*, 2019), which we also adopt.

### 2.2 Benchmarks and evaluation practice

V2X-Sim provides multi-agent LiDAR simulation with vehicles and a roadside unit (Li *et al.*, 2022), DAIR-V2X is the first large-scale real vehicle-infrastructure dataset (Yu *et al.*, 2022), and V2X-Real extends real collection to several vehicles and several infrastructure units (Xiang *et al.*, 2024). Published results on them come from heterogeneous protocols that differ in class definitions, evaluation ranges and data splits. Auditing evaluation practice is itself an established genre in transportation research: Parady *et al.* (2021) review validation practice across the discrete-choice literature and argue that goodness-of-fit reporting has displaced genuine validation. We are not aware of an equivalent audit for cooperative perception.

## 3. STUDY DESIGN

### 3.1 Benchmarks and data

Three benchmarks are used. (i) V2X-Sim (Li *et al.*, 2022), regenerated to restore the roadside unit that public re-splits omit, and evaluated on a held-out set of 2,500 frames. Absolute numbers are not comparable with the original release because of that re-split. (ii) DAIR-V2X (Yu *et al.*, 2022), real intersections observed by one vehicle and one infrastructure unit, evaluated for the car class under a scenario-level re-split that prevents scene leakage. Its absolute numbers are likewise not comparable with published DAIR-V2X results. (iii) V2X-Real (Xiang *et al.*, 2024) under the official vehicle-centric protocol: three classes (vehicle, pedestrian, truck), a longitudinal range of 100 m either side of the ego vehicle, and average precision (AP) at intersection-over-union thresholds of 0.3 and 0.5, to which we add 0.7 in our own runs. A detection counts as correct when its box overlaps a labelled object by at least the stated fraction, so a higher threshold tests localisation more strictly. Class-mean AP is written mAP, and every accuracy here is in percentage points.

### 3.2 Detection pipeline

All experiments use one pipeline family: a PointPillars encoder (Lang *et al.*, 2019) with V2X-ViT attention fusion (Xu *et al.*, 2022b). The V2X-Real experiments run in the official V2X-Real codebase, and the V2X-Sim and DAIR-V2X experiments run in a CoAlign-derived codebase (Lu *et al.*, 2023). Our V2X-Real runs use a 60-epoch, batch-size-2 schedule on the LiDAR-128 data variant instead of the published training configuration. Under the official protocol this reproduction reaches 40.3 mAP@0.5 against the published 35.8 for the same architecture, and every intermediate-fusion baseline we retrain matches or exceeds its published accuracy (Figure 2). The excess most plausibly reflects the longer schedule and higher-resolution sensor variant, not a faithful reproduction.

Two caveats bound these numbers. A second seed of the V2X-ViT reproduction gives 39.0 mAP@0.5, so the two-seed mean is 39.7 and run-to-run variation is of order one point. That variation exceeds the 0.1-point gap we measure between F-Cooper and attention fusion. Published-baseline validation also exists for V2X-Real only, because no published reference covers our V2X-Sim and DAIR-V2X re-splits.

### 3.3 Audit probes

We call any participant an agent and any non-ego agent a partner, and we apply five probes. (a) Agent ablation: evaluate one model with all agents, without infrastructure, and with the ego vehicle alone. (b) Pose-noise injection: add zero-mean Gaussian noise of increasing standard deviation to vehicle poses at test time. (c) Adversarial partner: corrupt one partner's transmitted features with additive zero-mean Gaussian noise at three times the standard deviation of that agent's own feature map. The corruption is untargeted and was not optimised against the detector. (d) Protocol variation: evaluate under an ad-hoc protocol that merges all classes into one and restricts evaluation to a symmetric 76.8 m window, then under the official multi-class protocol. (e) Stratification with scene control: among frames that have an infrastructure unit inside a 70 m communication range, separate those with no other connected vehicle in range (partner-poor) from those with at least one (partner-rich), and measure the marginal gain of the infrastructure unit in each stratum. The comparison runs on the validation split, on the broader test split, and inside one scenario containing both strata, so scene identity cannot masquerade as an availability effect.

## 4. FINDING 1: LITTLE RECOVERABLE HEADROOM IN SIMULATION

### 4.1 Infrastructure is redundant in aggregate, but not at range

Table 1 reports the agent-ablation controls on V2X-Sim. Cooperation as a whole is clearly valuable, since removing all partners costs 6.4 AP. Removing only the infrastructure unit costs 0.1 AP. With several connected vehicles present, the roadside viewpoint is close to redundant. A visibility analysis indicates why: regions observed only by infrastructure cover 23 to 25% of the monitored area but only 5.8% of the area occupied by labelled objects, because the simulator places vehicles densely enough that object locations are seen repeatedly.

Aggregate redundancy hides structure that a road authority would want. Beyond 30 m from the ego vehicle the infrastructure-exclusive gain rises from 0.1 to 1.2 AP. Even under dense collaboration the roadside viewpoint contributes specifically at range, which is where siting decisions bite, and the aggregate number conceals that entirely.

Table 1. Controls on V2X-Sim, one clean-trained model; change measured against row one.

| Configuration | AP@0.5 (%) | Change |
|---|---|---|
| All agents (vehicles and infrastructure) | 94.8 | reference |
| Without infrastructure | 94.7 | -0.1 |
| Ego vehicle only | 88.4 | -6.4 |
| One corrupted partner, no defence | 94.1 | -0.7 |
| Corrupted partner removed (oracle) | 93.9 | -0.9 |
| Trust gate, infrastructure-anchored / ego-anchored | 81.5 / 73.7 | -13.3 / -21.1 |

### 4.2 Pose error degrades accuracy, but leaves little for a remedy to recover

Figure 1 shows what happens when Gaussian noise is injected into vehicle poses at test time. Two models are compared: one trained with pose-noise augmentation, and a variant of it that re-estimates each vehicle's pose by registering its features against the infrastructure unit as a spatial anchor. This anchored correction is our own diagnostic probe, not a published method, and it is the natural way to exploit an infrastructure unit whose pose is exact.

Two readings must be kept apart. Accuracy itself is not saturated. The anchored variant falls from 91 to 82 AP as the noise standard deviation grows to 1.0 m, and the baseline falls from 90 to 85 over the 0 to 0.6 m range for which it was tabulated. Degradation on that scale is anything but negligible. What is saturated is the share of it that an anchored correction recovers: the gap holds at about one AP point at every level where both series were tabulated, and does not widen with the noise. Both models here are pose-noise-augmented, so their levels sit about 5 AP below the clean-trained diagnostic model of Table 1. The fixed gap is structural: because the simulated infrastructure pose is exact, the fusion backbone can already lean on that clean anchor implicitly, so an explicit correction has little left to recover. Remedies of this anchored family cannot demonstrate value here, though a mechanism of a different kind might.

### 4.3 The adversarial axis is equally closed

The adversarial controls in Table 1 close the second axis. Corrupting one partner's transmitted features costs 0.7 AP, because redundant viewpoints outvote it. Removing that partner outright gives 93.9 AP, no better than leaving it in, so an oracle defence has nothing to win either. Both trust gates then cost between 13 and 21 AP, since with no threat present they mostly suppress useful information from honest agents. One ordering does survive: a trust score computed against the infrastructure separates trustworthy from untrustworthy partners with an area under the receiver-operating characteristic of 0.785 and is unaffected when the ego pose degrades, where an ego-anchored score falls from 0.685 to 0.651. Under the single untargeted corruption we tested that ordering has no practical consequence. A corruption optimised against the detector might reopen the axis, and we have not tested one.

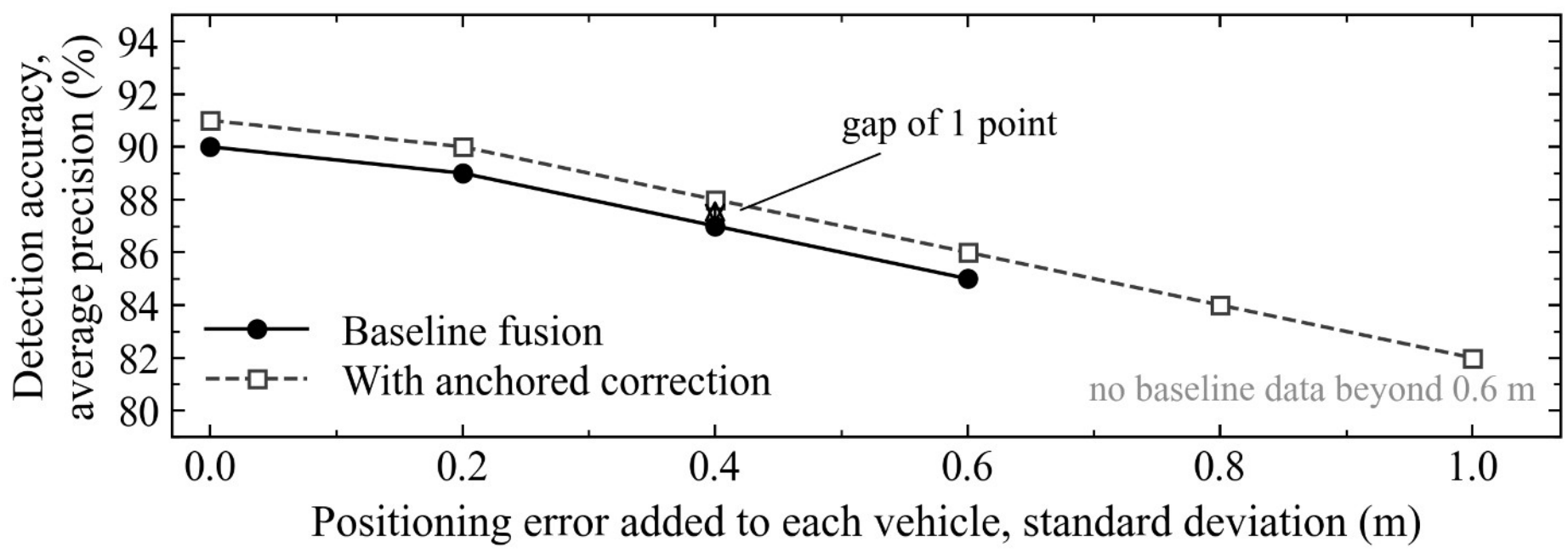

Figure 1. Detection accuracy on V2X-Sim as positioning error is added to each vehicle at test time. The gap between the two models stays near one point wherever both were evaluated.

### 4.4 Mechanism and implication

Both closures share one candidate mechanism. Dense partner placement duplicates viewpoints, and exact infrastructure poses remove the nuisance that robustness methods target. We did not test that mechanism directly. The decisive control would perturb the infrastructure pose and show the axis reopening, whereas our simulation experiments all keep the roadside pose clean. The attribution therefore rests on how the simulator is built and not on a measurement, and we leave the control to future work. One reading rule still follows. Neither condition holds on real roads, where penetration is around 6% (Kandiboina *et al.*, 2024) and roadside units are calibrated imperfectly, so robustness results from this setting are upper bounds under best-case assumptions and not evidence of deployment readiness.

## 5. FINDING 2: MEASUREMENT CHOICES GOVERN THE CONCLUSION

### 5.1 Class definition and evaluation range

Our own first protocol for V2X-Real was ad hoc. All classes were merged into one, and evaluation was restricted to a symmetric 76.8 m window instead of the official 100 m. Measured that way, full cooperation improves detection over the ego-only configuration by 1.6 AP, from 19.4 to 21.0. To anyone using this evidence that margin would read as a weak case for roadside deployment. Under the official three-class protocol on the same dataset, the published benefit of cooperative fusion is 14.6 mAP, from 21.2 for no fusion to 35.8 for V2X-ViT fusion (Xiang *et al.*, 2024).

These two figures are not a controlled protocol comparison and we do not present them as one. Our ad-hoc measurement used a subset-trained model, while the published figures come from fully trained ones, so the contrast bounds the protocol effect without isolating it. The control that would isolate it is an ego-only evaluation of our own model under the official protocol, which we did not run. That omission is worth naming, since it is exactly the negative control Section 6.3 asks of others. Measurement does support the mechanism. Under the official protocol our reproduced V2X-ViT detects vehicles at 64.8 AP@0.5, pedestrians at 23.4 and trucks at 32.7. Merging everything into one vehicle-dominated class therefore dilutes the two classes on which a second viewpoint helps most, and the narrowed window discards far objects that only partners observe. Consistent with that account, moving our pipeline to the official protocol without touching the architecture revealed cooperative headroom of the published size. Figure 2 sets our retrained baselines beside the published values: the intermediate-fusion reproductions reach 36.9 to 40.3 mAP against a published no-fusion level of 21.2.

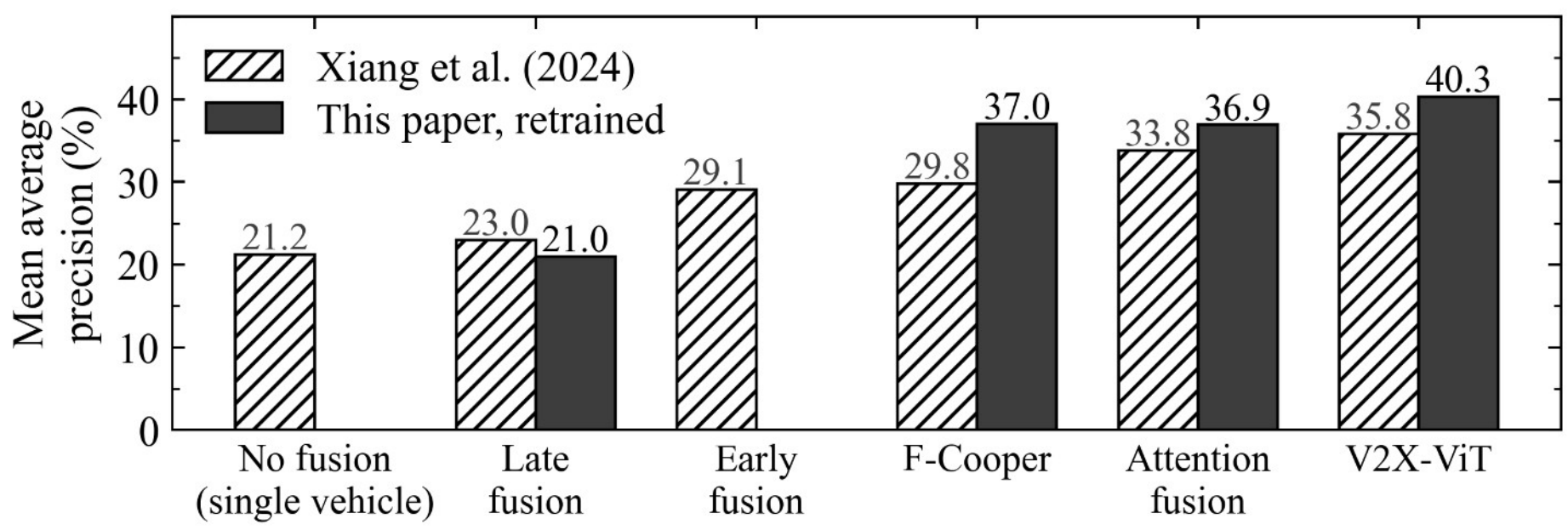

Figure 2. Class-mean accuracy on the official V2X-Real benchmark: values reported by Xiang *et al.* (2024) beside our retrained single-seed baselines. No fusion and early fusion were not retrained.

### 5.2 Scene composition masquerades as a partner-availability effect

A policy-relevant question is whether roadside infrastructure helps most when few connected vehicles are nearby, since that is the regime early deployments will face (Kandiboina *et al.*, 2024). We test it with the stratification of Section 3.3. Among frames with an infrastructure unit in range, we compare the marginal gain of that unit in frames with no other connected vehicle against frames with at least one. A naive analysis appears to confirm the hypothesis dramatically. On the validation split the infrastructure is worth 3.4 AP in the 123 partner-poor frames and 0.2 AP in the 185 partner-rich ones, a ratio of roughly seventeen. That stratification is confounded twice over. Partner-poor frames contain systematically farther objects, at a mean distance of 41.9 m against 29.7 m, and with 37% of objects beyond 50 m against 11%. No validation scenario contributes frames to both strata either, so scene identity and partner availability cannot be separated at all.

Table 2 shows what happens as the analysis is repaired, and it carries a second lesson we did not expect. Moving to the test split, which spans more scenes, takes the ratio of absolute gains from seventeen to 3.3. Restricting the comparison to the one test scenario that contributes frames to both strata takes it to 1.7. The seventeen-fold figure does not survive: no reading of the test split exceeds four-fold. Which way the scene control itself moves the ratio, however, depends on the measurement basis. On absolute gains the subset-trained model falls from 3.3 to 1.7. On relative gains the same comparison rises, from 1.7 to 2.1. The fully trained model falls on the relative basis, from 3.9 to 1.6. From these same runs, a paper reporting only one basis could claim either that scene matching removes most of the effect or that it leaves the effect intact. Three caveats belong with these numbers. All were measured under the ad-hoc protocol of Section 5.1, so they size the stratification artefact and not the deployable benefit of infrastructure. The naive ratio is unstable because its denominator, 0.2 AP, is near zero. The scene-matched comparison rests on one scenario contributing fewer than 80 frames per stratum. What is unambiguous is the negative part: the seventeen-fold claim is an artefact of how frames were grouped.

Table 2. Marginal benefit of the infrastructure unit by partner availability, under the ad-hoc protocol of Section 5.1; (subset) and (full) mark the subset-trained and fully trained model. Gains are AP points, relative gain in brackets, then the poor-to-rich ratio on each basis.

| Analysis | Poor | Rich | Abs. | Rel. |
|---|---|---|---|---|
| Validation, naive (subset) | +3.4 (+47%) | +0.2 | ~17x | n/a |
| Test, naive (subset) | +5.9 (+17.5%) | +1.8 (+10.5%) | 3.3x | 1.7x |
| Test, scene-matched (subset) | +4.0 (+23%) | +2.3 (+11%) | 1.7x | 2.1x |
| Test, naive (full) | +3.8 (+7.1%) | +0.5 (+1.8%) | 7.6x | 3.9x |
| Test, scene-matched (full) | +12.9% | +8.3% | n/a | 1.6x |

## 6. FINDING 3: REAL-WORLD VALUE AND VOLATILE COLLABORATION

### 6.1 Cooperation pays on real data

The audited protocols leave the central promise of cooperative perception intact. On DAIR-V2X, adding the single infrastructure viewpoint raises car detection from 68.5 to 75.6 AP@0.5 under our scenario-level re-split, a gain of 7.1 points. On V2X-Real under the official protocol, cooperative fusion adds 14.6 mAP over single-vehicle perception (Xiang *et al.*, 2024), and our retrained intermediate-fusion baselines all match or exceed their published accuracy, though the order of two adjacent ones is not preserved and their gap lies inside our seed-to-seed variation (Section 3.2). For a road authority the useful reading is directional: one roadside unit at an intersection is worth several accuracy points to a vehicle approaching it, which is the quantity a siting model such as that of Zeng and He (2024) would consume.

### 6.2 Partner availability is volatile, and split-dependent

Figure 3 shows how many partners are actually within communication range of the ego vehicle on the V2X-Real validation split, counting agents inside a 70 m range over 717 frames. The count varies from zero to three. In 27% of frames the ego vehicle has no partner at all, and only 43% of frames have an infrastructure unit in range. A single aggregate accuracy number therefore averages over materially different operating conditions.

The picture changes completely on the other split. A stratification of the test data under the official protocol, grouping frames by partners within communication range, leaves only 7 of the 1,253 frames in the nine test scenarios that pipeline retains with no partner, 351 with one and 895 with two or more. The two counts come from different runs, and the communication range of the test run is not separately recorded, so split composition cannot be fully separated from construction detail. A regime-conditioned claim inherits both the split and the construction that produced it, and it must state them. On those nine scenarios, conditioning accuracy on availability gives a real but moderate difference: 38.5 mAP@0.5 with one partner against 41.3 with two or more.

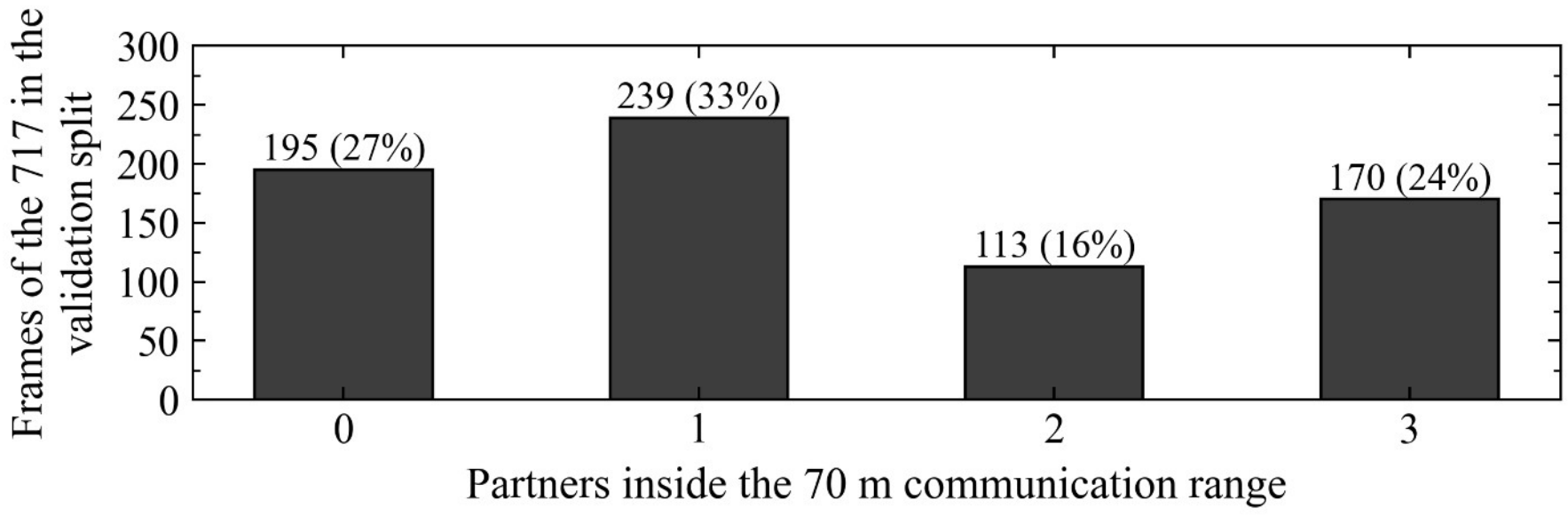

Figure 3. Partners inside a 70 m communication range, counted per frame over the 717 frames of the V2X-Real validation split.

### 6.3 Implications for evaluation practice

Five practices follow from the audit. Evaluate under each benchmark's official multi-class protocol and report per-class accuracy, since collapsing classes and narrowing the range remove the very distances and classes at which a second viewpoint contributes. Report negative controls, meaning ego-only and no-infrastructure ablations, so that the marginal value of each cooperation channel is visible and not merely implied; Section 5.1 shows what a missing one costs. Stratify by partner availability, name the split, control scene composition before attributing differences to traffic conditions, and state the surviving sample size. Treat robustness and security results from dense, clean simulation as upper bounds, and model infrastructure pose error and low penetration before drawing deployment conclusions. Retrain published baselines under a stated schedule and report both numbers with run-to-run variation: ours departed from published accuracy by between -2.0 and +7.2 mAP, and part of that spread comes from our longer schedule and higher-resolution sensor variant, not from reproduction noise.

## 7. CONCLUSION

This paper audited the evidence base of cooperative perception across one simulated and two real-world V2X benchmarks with a single pipeline family. In simulation, dense partner placement and exact infrastructure poses leave almost nothing for an anchored pose correction or a trust gate to recover: the correction returns about one accuracy point at every error level, corrupting a partner costs 0.7 points, and removing that partner outright is no better than leaving it in. On real data, measurement choices govern the conclusion. An apparent seventeen-fold advantage of infrastructure in partner-poor frames falls below four-fold once the split is broadened, and scene matching then moves it in opposite directions depending on whether gains are counted in absolute or relative terms; an ad-hoc class definition measures a far smaller cooperation benefit than the official protocol reports. Cooperation itself remains worth 7 to 15 accuracy points on real data, yet partner availability differs so sharply between splits, from 27% partner-free frames on one to almost none on the other, that every regime-conditioned claim must name the split and the construction behind it. Extending the audit to communication latency, packet loss and bandwidth limits, which dominate real V2X links, is the natural next step.